\documentstyle[12pt,epsf]{article}
\begin{document}
\begin{center}
{\Large \bf Low-temperature expansion and perturbation theory 
in $2D$ models with unbroken symmetry: a new approach }\\
\vspace*{1cm}
{\bf O.~Borisenko\footnote{email: oleg@ap3.bitp.kiev.ua}, \
V.~Kushnir\footnote{email: kushnir@ap3.bitp.kiev.ua}, \
A.~Velytsky\footnote{email: vel@ap3.bitp.kiev.ua}}\\

\vspace*{0.5cm}
{\large \it
N.N.Bogolyubov Institute for Theoretical Physics, National Academy
of Sciences of Ukraine, 252143 Kiev, Ukraine}\\
\vspace*{0.3cm}

\end{center}
\vspace{.5cm}

\begin{abstract}
A new method of constructing a weak coupling expansion of two 
dimensional ($2D$) models with an unbroken continuous symmetry is 
developed. The method is based on an analogy with the abelian $XY$ model,
respects the Mermin-Wagner (MW) theorem and uses a link representation 
of the partition and correlation functions. An expansion 
of the free energy and of the correlation functions at small 
temperatures is performed and first order coefficients
are calculated explicitly. They are shown to coincide with 
the results of the conventional perturbation theory.
We discuss an applicability of our method to analysis
of uniformity of the low-temperature expansion.
\end{abstract}

\section{Introduction}

Since two-dimensional models with a continuous global symmetry
were recognized to be asymptotically free \cite{free}
they became a famous laboratory for testing many ideas
and methods before applying them to more complicated gauge theories.
In this paper we follow this common way and present a method,
different from the conventional perturbation theory (PT), 
which allows to investigate
these models in the weak coupling region.
The conventional PT is one of the main technical
tools of the modern physics. In spite of the belief of the most
of community that this method gives the correct asymptotic expansion
of such theories as $4D$ QCD or $2D$ spin systems with continuous global
symmetry, the recent discussion of this problem
\cite{superinst}-\cite{QA} has shown that it is rather far from
unambiguous solution. 
Indeed, for the PT to be applicable it is necessary
that a system under consideration would possess a well ordered
ground state. In two dimensional models, like the $O(N)$-sigma models
the MW theorem guarantees the absence of such a state in 
the thermodynamic limit (TL) however small coupling constant is \cite{MW}.
Then, the usual argument in support of the PT is that locally
the system is ordered and the PT is not supposed to be used for the calculation
of long-distance observables. On the other hand it should reproduce the correct
behaviour of the fixed-distance correlations as well as
all thermodynamical functions which can be expressed via
short-range observables. The example of $1D$ models shows that
even this is not always true \cite{rossi}, 
so why should one believe in correctness of the conventional PT in $2D$?
In fact, the only way to justify the PT is to
prove that it gives the correct asymptotic expansion of 
nonperturbatively defined models in the TL.
Now, it was shown in \cite{superinst}
that PT results in $2D$ nonabelian models depend on the boundary
conditions (BC) used to reach the TL.
This result could potentially imply that the low-temperature limit and
the TL do not commute in nonabelian models. Actually, the main 
argumentation of \cite{superinst} regarding the failure
of the PT expansion is based
on the fact that the conventional PT is an expansion around
a broken vacuum, i.e. the state which simply does not exist in
the TL of $2D$ models. According to \cite{superinst},
the ground state of these systems can be described
through special configurations -- the so-called gas of
superinstantons (SI) and the correct expansion should take into account
these saddle points. 
At the present stage it is rather unclear how one could
construct an expansion in the SI background. Fortunately, there exists
other, more eligible way to construct the low-temperature expansion
which respects the MW theorem and is apriori not the expansion around 
the broken vacuum. We develop this method on an example of
the $2D$ $SU(N)\times SU(N)$ principal chiral model
whose partition function (PF) is given by
\begin{equation}
Z = \int\prod_xDU_x
\exp \left [\beta \sum_{x,n} {\mbox {Re Tr}} U_xU_{x+n}^+\right] ,
\label{pfsun}
\end{equation}
\noindent
where $U_x\in SU(N)$, $DU_x$ is the invariant measure and we
impose the periodic BC. The basic idea is the following.
As was rigorously proven, the conventional PT 
gives an asymptotic expansion which is uniform
in the volume for the abelian $XY$ model \cite{XYPT}. 
One of the basic theorems which underlies the proof states 
that the following inequality holds in the $3D$ $XY$ model
\begin{equation}
< \exp (\sqrt{\beta}A(\phi_x)) > \ \leq \ C \ ,
\label{ineq3D}
\end{equation}
\noindent
where $C$ is $\beta$-independent and $A(\phi +2\pi)=\phi$. $\phi_x$
is an angle parametrizing the action of the $XY$ model
$S=\sum_{x,n}\cos (\phi_x - \phi_{x+n})$. It follows that at large
$\beta$ the Gibbs measure is concentrated around $\phi_x\approx 0$ 
providing a possibility to construct an expansion around $\phi_x=0$.
This inequality is not true in $2D$ in the thermodynamic limit
because of the MW theorem, however the authors of \cite{XYPT} prove 
the same inequality for the link angle, i.e.
\begin{equation}
< \exp (\sqrt{\beta}A(\phi_l)) > \ \leq \ C \ , \ 
\phi_l=\phi_x-\phi_{x+n} \ ,
\label{ineq2D}
\end{equation}
\noindent
where the expectation value refers to infinite volume limit.
Thus, in $2D$ the Gibbs measure at large $\beta$ is concentrated
around $\phi_l\approx 0$ and the asymptotic series can be
constructed expanding the action in powers of $\phi_l$. In the abelian
case such an expansion is equivalent to the expansion around $\phi_x=0$
because i) the action depends only on the difference $\phi_x - \phi_{x+n}$
and ii) the integration measure is flat, $DU_x=d\phi_x$. 

In $2D$ nonabelian models, again because of the MW theorem, one has to expect 
in the TL something alike to (\ref{ineq2D}), namely
\begin{equation}
< \exp (\sqrt{\beta}{\mbox {arg}}A({\mbox {Tr}}V_l)) >
\ \leq \ C \ , \ V_l=U_xU_{x+n}^+ \ .
\label{nabineq2D}
\end{equation}
\noindent
Despite there is not a rigorous proof of (\ref{nabineq2D}), 
that such (or similar) inequality holds in $2D$ nonabelian models 
is intuitively clear and should follow from the chessboard 
estimates \cite{chest} and from the Dobrushin-Shlosman proof of the MW theorem
\cite{MW} which shows that spin configurations are distributed uniformly
in the group space in the TL. Namely, the probability $p(\xi )$ 
that ${\mbox {Tr}}(V_l-I)\leq -\xi$ is bounded by 
\begin{equation}
p(\xi ) \leq O(e^{-b\beta\xi}) \ , \ \beta\to\infty \ ,
\label{chestlink}
\end{equation}
\noindent
if the volume is sufficiently large, $b$ is a constant. 
Thus, until $\xi\leq O((\sqrt{\beta})^{-1})$ is not satisfied,
all configurations are exponentially suppressed.
This is equivalent to the statement that the Gibbs 
measure at large $\beta$ is concentrated around $V_l\approx I$, therefore
(\ref{nabineq2D}) or its analog holds. 
In what follows it is assumed that (\ref{nabineq2D}) is correct, 
hence $V_l=I$ is the only saddle point for the invariant 
integrals\footnote{It follows already from (\ref{chestlink}).
What is important in (\ref{nabineq2D}) is a factor $\sqrt{\beta}$,
otherwise the very possibility of the expansion in $1/\beta$
becomes problematic.}. 
Thus, the correct asymptotic expansion, if exists,
should be given via an expansion around $V_l=I$, similarly to the
abelian model. If the conventional PT gives the correct asymptotics, it
must reproduce the series obtained expanding around $V_l=I$.
However, neither i) nor ii) holds in the nonabelian models, therefore
it is far from obvious that two expansions indeed coincide.
Let us parametrize $V_l=\exp (i\omega_l)$ and $U_x=\exp (i\omega_x)$.
Consider the following expansion
\begin{equation}
V_l = \exp \left [i\omega_l \right ] \sim 
I + \sum_{n=1}\frac{1}{(\beta)^{n/2}}\frac{(i\omega_l)^n}{n!} \ .
\label{ltexp}
\end{equation}
\noindent
The standard PT states that to calculate the asymptotic expansion
one has to re-expand this series as
\begin{equation}
\omega_l = \omega_x-\omega_{x+n}+ \sum_{k=1}
\frac{1}{(\beta)^{k/2}}\omega_l^{(k)} \ ,
\label{ltosres}
\end{equation}
\noindent
where $\omega_l^{(k)}$ are to be calculated from the definition
$U_xU_{x+n}^+=\exp (i\omega_l)$.
This is presumably true in a finite volume where one can fix
appropriate BC (like the Dirichlet ones), or to break down 
the global symmetry by fixing the global gauge 
(on the periodic lattice). Then, making $\beta$ sufficiently large
one forces all the spin matrices to fluctuate around $U_x\approx I$,
therefore the substitution (\ref{ltosres}) is justified.
We do not see how this procedure could be justified when
the volume increases and fluctuations of $U_x$ spread up
over the whole group space. In other words, it is not clear why
in the series
$$
\omega_l = \omega_x-\omega_{x+n}+ \sum_{k=1}\omega_l^{(k)} \ 
$$
the term $\omega_l^{(k+1)}$ is suppressed as $\beta^{-1/2}$
relatively to the term  $\omega_l^{(k)}$.
It is only (\ref{ltexp}) which remains correct
in the large volume limit and takes into account 
all the fluctuations contributing at a given order of 
the low-temperature expansion.

It is a purpose of the present paper to develop
an expansion around $V_l=I$ aiming to calculate the asymptotic
series for nonabelian models.
First of all, one has to give a precise mathematical meaning to
the expansion (\ref{ltexp}). It is done in the next section. 

\section{Link representation for the partition and correlation functions}

To construct an expansion of the Gibbs measure 
and the correlation functions using (\ref{ltexp}) 
we use the so-called link representation for the partition and 
correlation functions. First, we make a change of variables
$V_l=U_xU_{x+n}^+$ in (\ref{pfsun}). PF becomes
\begin{equation}
Z = \int \prod_l dV_l
\exp \left[ \beta \sum_l {\mbox {Re Tr}} V_l + \ln J(V) \right] \ ,
\label{lPF}
\end{equation}
\noindent
where the Jacobian $J(V)$ is given by \cite{linkrepr}
\begin{equation}
J(V) = \int \prod_xdU_x\prod_l
\left[ \sum_r d_r \chi_r \left( V^+_lU_xU^+_{x+n} \right) \right]
= \prod_p \left[ \sum_r d_r \chi_r \left( \prod_{l\in p}V_l \right) \right].
\label{jacob}
\end{equation}
\noindent
$\prod_p$ is a product over all plaquettes of $2D$ lattice,
the sum over $r$ is sum over all representations of $SU(N)$, 
$d_r=\chi_r(I)$ is the dimension of $r$-th representation. 
The $SU(N)$ character $\chi_r$ depends on a product of the link 
matrices $V_l$ along a closed path (plaquette in our case):
\begin{equation}
\prod_{l\in p}V_l = V_n(x)V_m(x+n)V_n^+(x+m)V_m^+(x) \ .
\label{prod}
\end{equation}
\noindent
The expression $\sum_r d_r \chi _r(\prod_{l\in p}V_l)$ is the
$SU(N)$ delta-function which reflects the fact that the product of
$U_xU^+_{x+n}$ around plaquette equals $I$ (original model has
$L^2$ degrees of freedom, $L^2$ is a number of sites; since a number
of links on the $2D$ periodic lattice is $2L^2$, the Jacobian must
generate $L^2$ constraints)\footnote{Strictly speaking, on the periodic
lattice one has to constraint two holonomy operators, i.e. closed
paths winding around the whole lattice. We do not expect such global 
constraints to influence the TL in $2D$ (see Discussion).}. 
The solution of the constraint
\begin{equation}
\prod_{l\in p}V_l = I
\label{constr}
\end{equation}
\noindent
is a pure gauge $V_l=U_xU_{x+n}^+$, so that two forms of the PF
are exactly equivalent.

The corresponding representation for the abelian $XY$ model reads
\begin{equation}
Z_{XY} = \int\prod_ld\phi_l
\exp \left [\beta \sum_l\cos\phi_l \right ]\prod_pJ_p \ ,
\label{PFLxy}
\end{equation}
\noindent
where the Jacobian is given by the periodic delta-function
\begin{equation}
J_p = \sum_{r=-\infty}^{\infty} e^{ir\phi_p} \ , \
\phi_p=\phi_n(x)+\phi_m(x+n)-\phi_n(x+m)-\phi_m(x+n) \ .
\label{PFLxyJ}
\end{equation}
\noindent

Consider two-point correlation function
\begin{equation}
\Gamma (x,y) = < {\mbox {Tr}} \ U_xU_y^+ > \ ,
\label{corf}
\end{equation}
\noindent
where the expectation value refers to the ensemble defined in (\ref{pfsun}). 
Let $C_{xy}$ be some path connecting points $x$ and $y$. Inserting 
the unity $U_zU_z^+$ in every site $z\in C_{xy}$ one gets
\begin{equation}
\Gamma (x,y) = < {\mbox {Tr}} \prod_{l\in C_{xy}} (U_xU_{x+n}^+) > =
< {\mbox {Tr}} \prod_{l\in C_{xy}} W_l > \ ,
\label{corf1}
\end{equation}
\noindent
where $W_l = V_l$ if along the path $C_{xy}$ the link $l$ goes in
the positive direction and $W_l = V_l^+$, otherwise.  
The expectation value in (\ref{corf1}) refers now to the ensemble
defined in (\ref{lPF}). Obviously, it does not depend on the path $C_{xy}$
which can be deformed for example to the shortest path between sites $x$ 
and $y$.

In this representation the series (\ref{ltexp}) acquires a well
defined meaning, therefore the expansion of the action,
of the invariant measure, etc. can be done.

\section{$XY$ model: Weak coupling expansion of the free energy}

In this section we prove that for the abelian XY model 
the large-$\beta$ expansion in the link
representation gives the same results as the conventional PT 
in the thermodynamic limit. We consider only the free energy but 
the generalization for the correlation functions is straightforward.

The first step is a standard one, i.e. we rescale 
$\phi\to\frac{\phi}{\sqrt{\beta}}$ and make an expansion 
\begin{equation}
\exp \left[ \beta\cos\frac{\phi}{\sqrt{\beta}} \right] = \exp 
\left[ \beta -\frac{1}{2}(\phi)^2 \right] 
\left[ 1 + \sum_{k=1}^{\infty} (\beta)^{-k} 
\sum_{l_1,..,l_k}\frac{a_1^{l_1}...a_k^{l_k}}{l_1!...l_k!} \right] \ ,
\label{cos}
\end{equation}
\noindent  
where $l_1+2l_2+...+kl_k=k$ and
\begin{equation}
a_k = (-1)^{k+1}\frac{\phi^{2(k+1)}}{(2k+2)!} \ .
\label{a_k}
\end{equation}
\noindent
In addition to this perturbation one has to extend the integration region
to infinity. We do not treat this second perturbation, as usually supposing 
that all the corrections from this perturbation go down exponentially with 
$\beta$ (in the abelian case it can be proven rigorously \cite{XYPT}). 
It is more convenient now to go to a dual lattice identifying 
plaquettes of the original lattice with its center, i.e. $p\to x$.
Let $l=(x;n)$ be a link on the dual lattice.
Introducing sources $h_l$ for the link field, one then finds 
\begin{equation}
Z_{XY}(\beta >> 1) = e^{\beta DL^2 - L^2\ln \beta}
\prod_{l} \left[ 1+\sum_{k=1}^{\infty}\frac{1}{\beta^k}
A_k\left( \frac{\partial^2}{\partial h_l^2} \right) \right] M(h_l) \ .
\label{asxyGll}
\end{equation}
\noindent
Coefficients $A_k$ are defined in (\ref{cos}) and (\ref{a_k}).
The generating functional $M(h_l)$ is given by
\begin{eqnarray}
M(h_l) = \sum_{r_x=-\infty}^{\infty}\int_{-\infty}^{\infty}
\prod_ld\phi_l \exp 
\left[-\frac{1}{2}\sum_l \phi_l^2 
+ i\sum_l \frac{\phi_l}{\sqrt{\beta}} (r_x-r_{x+n}) 
+\sum_l\phi_l h_l  \right] .
\label{Gfunxy}
\end{eqnarray}
\noindent
Sums over representations are treated by the Poisson resummation formula.
The Gaussian ensemble appears in terms of the fluctuations of $r$-fields.
The integral over zero mode of $r$-field is not Gaussian and leads to 
a delta-function in the Poisson formula. Thus, the zero mode 
decouples from the expansion.
All the corrections to the integrals over representations in the
Poisson formula fall down exponentially with $\beta$ so that 
the generating functional becomes
\begin{equation}
M(h_l) = \exp 
\left[ \frac{1}{4} \sum_{l,l^{\prime}}h_lG_{ll^{\prime}}h_{l^{\prime}} \right] \ ,
\label{Gfunxy1}
\end{equation}
\noindent
where we have introduced the following function which we term 
``link'' Green function 
\begin{equation}
G_{ll^{\prime}} = 2\delta_{l,l^{\prime}} - G_{x,x^{\prime}} -
G_{x+n,x^{\prime}+n^{\prime}} + G_{x,x^{\prime}+n^{\prime}} + G_{x+n,x^{\prime}} \ .
\label{Gll1}
\end{equation}
\noindent
$G_{x,x^{\prime}}$ is a ``standard'' Green function on the periodic lattice
\begin{equation}
G_{x,x^{\prime}} = \frac{1}{L^2} \sum_{k_n=0}^{L-1}
\frac{e^{\frac{2\pi i}{L}k_n(x-x^{\prime})_n}}
{D-\sum_{n=1}^D\cos \frac{2\pi}{L}k_n} \ , \ k_n^2\ne 0 \ .
\label{Gxx}
\end{equation}
\noindent
Normalization is such that $G_{ll}=1$.
As far as we could check, the expression (\ref{asxyGll}) reproduces  
the well known asymptotics of the free energy of the $XY$ model 
in two dimensions. For example, the first order coefficient of the 
free energy being expressed in $G_{ll^{\prime}}$ reads
\begin{equation}
C_1 = \frac{1}{64 L^2} \sum_{l} G^2_{ll} = \frac{1}{32} \ .
\label{C1Gll}
\end{equation}
\noindent
Let us add some comments. In the standard expansion to avoid the zero mode problem
one has to fix appropriate BC, like Dirichlet ones or to fix a global gauge
if one works on the periodic lattice \cite{unitgauge}. In the present scheme
the zero mode decouples automatically due to using $U(1)$ delta-function
which takes into account the periodicity of the integrand in link angles.
More important observation is that it is allowed
to take the TL already in the formula (\ref{asxyGll}):
since the generating functional
depends only on the link Green function which is infrared finite, 
the uniformity of the expansion in the volume follows immediately.  
This is a direct consequence of the fact that the Gibbs measure of 
the $XY$ model is a function of the gradient $\phi_l$ only.

\section{Weak coupling expansion in the $SU(2)$ model}

We turn now to the nonabelian models. As the simplest example we analyze
the $SU(2)$ principal chiral model. The method developed here 
has straightforward generalization to arbitrary $SU(N)$ or $SO(N)$ model 
and we shall present it elsewhere. 

\subsection{Representation for the partition function}

We take the standard form for the $SU(2)$ link matrix which is the most suitable
for the weak coupling expansion
\begin{equation}
V_l = \exp [i\sigma^k\omega_k(l)] \ ,
\label{su2mtr}
\end{equation}
\noindent
where $\sigma^k, k=1,2,3$ are Pauli matrices.
Let us define
\begin{equation}
W_l = \left[ \sum_k\omega^2_k(l) \right]^{1/2}
\label{Wl}
\end{equation}
\noindent
and similarly
\begin{equation}
W_p = \left[ \sum_k\omega^2_k(p) \right]^{1/2} ,
\label{Wp}
\end{equation}
\noindent
where $\omega_k(p)$ is a plaquette angle defined as
\begin{equation}
V_p = \prod_{l\in p}V_l =  \exp \left [ i\sigma^k\omega_k(p) \right ] \ .
\label{plangle}
\end{equation}
\noindent
The exact relation between link angles $\omega_k(l)$ and the plaquette
angle $\omega_k(p)$ is given in the Appendix B. Then, the partition function
(\ref{lPF}) can be exactly rewritten to the following form appropriate
for the weak coupling expansion
\begin{eqnarray}
Z = \int \prod_l \left[ \frac{\sin^2W_l}{W^2_l}\prod_kd\omega_k(l) \right]
\exp \left[ 2\beta\sum_l\cos W_l \right] \prod_x \frac{W_x}{\sin W_x} 
\nonumber     \\
\prod_x \sum_{m(x)=-\infty}^{\infty}\int\prod_kd\alpha_k(x)
\exp \left[ -i\sum_k\alpha_k(x)\omega_k(x) + 2\pi im(x)\alpha (x) \right] \ ,
\label{PFwk}
\end{eqnarray}
\noindent
where we have introduced auxiliary field $\alpha_k(x)$ and
\begin{equation}
\alpha (x) = \left[ \sum_k\alpha^2_k(x) \right]^{1/2} .
\label{Ax}
\end{equation}
\noindent
The representation for the Jacobian (\ref{D14}) used here can be 
interpreted as an analog of the $U(1)$ periodic delta-function 
(\ref{PFLxyJ}): this is the $SU(2)$ delta-function which is periodic
with respect to a length of the vector $\vec{\omega}(x)$.
We give a derivation of the partition function (\ref{PFwk})
in the Appendix A. In rewriting the final formula of that derivation
(\ref{D19}), we went over to the dual lattice identifying 
the links of the original lattice with dual links and the original
plaquettes with dual sites located in the center of the original
plaquettes.

\subsection{General expansion}

To perform the weak coupling expansion 
we proceed in a standard way, i.e. first we make the substitution
\begin{equation}
\omega_k(l)\to (2\beta)^{-1/2}\omega_k(l) \ , \
\alpha_k(x)\to (2\beta)^{1/2}\alpha_k(x) 
\label{subst}
\end{equation}
\noindent
and then expand the integrand of (\ref{PFwk}) 
in powers of fluctuations of the link fields.
We would like to give here some technical details of the expansion
which could be useful for a future use. 
It is straightforward to get the following power series:
\begin{enumerate}

\item Action

\begin{equation}
\exp \left[ 2\beta\cos\frac{W_l}{\sqrt{2\beta}} \right] = \exp 
\left[ 2\beta -\frac{1}{2}(W_l)^2 \right] 
\left[ 1 + \sum_{k=1}^{\infty} (2\beta )^{-k} 
\sum_{l_1,..,l_k} \frac{a_1^{l_1}...a_k^{l_k}}{l_1!...l_k!} \right] \ ,
\label{actexp}
\end{equation}
\noindent  
where $l_1+2l_2+...+kl_k=k$ and
\begin{equation}
a_k = (-1)^{k+1}\frac{W_l^{2(k+1)}}{(2k+2)!} \ .
\label{a_kW}
\end{equation}
\noindent

\item Invariant measure

\begin{equation}
\frac{\sin^2W_l}{W^2_l} = 1 +
\sum_{k=1}^{\infty}\frac{(-1)^k}{(2\beta )^k}C_kW_l^{2k} \ , \ 
C_k = \sum_{n=0}^k \frac{1}{(2n+1)!(2k-2n+1)!} \ .
\label{invMexp}
\end{equation}
\noindent

\item Contribution from Jacobian I

\begin{equation}
\frac{W_x}{\sin W_x} = 
1 + \sum_{k=1}^{\infty}\frac{J_k}{(2\beta )^k}W_x^{2k} \ , \
J_k = 2\frac{2^{2k-1}}{(2k)!}\mid B_{2k} \mid \ ,
\label{J1exp}
\end{equation}
\noindent
where $B_{2k}$ are Bernoulli numbers.

\item Contribution from Jacobian II

\begin{eqnarray}
\alpha_k(x)\omega_k(x) = \alpha_k(x)\left[ \omega^{(0)}_k(x) +
\sum_{n=1}^{\infty} \frac{\omega^{(n)}_k(x)}{(2\beta )^{n/2}} \right] 
\ , \\ \nonumber
\exp \left[ -i\sum_k\alpha_k(x)\omega_k(x) \right] = 
\exp \left[ -i\sum_k\alpha_k(x)\omega^{(0)}_k(x) \right] \\  \nonumber
\left[ 1 + \sum_{q=1}^{\infty}\frac{(-i)^q}{q!} \left( \sum_k\alpha_k(x) 
\sum_{n=1}^{\infty} \frac{\omega^{(n)}_k(x)}{(2\beta )^{n/2}} \right)^q \right] \ .
\label{J2exp}
\end{eqnarray}
\noindent

\end{enumerate}
Using the relations (\ref{D9})-(\ref{D12}) one can calculate 
$\omega^{(n)}_k(x)$ up to an arbitrary order in $n$. In particular, 
$\omega^{(0)}_k(x)$ is given by (see Fig.1 for our notations of
dual links)
\begin{equation}
\omega^{(0)}_k(x) = \omega_k(l_3) + \omega_k(l_4) - 
\omega_k(l_1) - \omega_k(l_2) \ .
\label{w0}
\end{equation}
\noindent
We shall use an obvious property
\begin{equation}
\sum_x\alpha_k(x)\omega^{(0)}_k(x) = 
\sum_l\omega_k(l) \left [\alpha_k(x+n) - \alpha_k(x) \right ] \ , \ l=(x;n) \ .
\label{resum}
\end{equation}
\noindent
Introducing now the external sources $h_k(l)$ coupled to the link field 
$\omega_k(l)$ and $s_k(x)$ coupled to the auxiliary field $\alpha_k(x)$
and adjusting the definitions
\begin{equation}
\omega_k(l) \to \frac{\partial}{\partial h_k(l)} \ , \ 
\alpha_k(x) \to \frac{\partial}{\partial s_k(x)} \ ,
\label{deriv}
\end{equation}
\noindent
we get finally the following formal weak coupling expansion 
for the PF (\ref{PFwk})
\begin{eqnarray}
Z = C(2\beta ) Z(0,0) \prod_l\left[ \left( 1 + \sum_{k=1}^{\infty} (2\beta )^{-k} 
\sum_{l_1,..,l_k}^{\prime}\frac{a_1^{l_1}...a_k^{l_k}}{l_1!...l_k!} \right) 
\left( 1 + \sum_{k=1}^{\infty}\frac{(-1)^k}{(2\beta )^k}C_kW_l^{2k} 
\right) \right]  \\  \nonumber
\prod_x \left[ \left(1 + \sum_{k=1}^{\infty}\frac{J_k}{(2\beta )^k}W_x^{2k} \right) 
\left( 1 + \sum_{q=1}^{\infty}\frac{(-i)^q}{q!} \left( \sum_k\alpha_k(x) 
\sum_{n=1}^{\infty} \frac{\omega^{(n)}_k(x)}{(2\beta )^{n/2}} \right)^q \right) \right]
\ M(h,s) \ ,
\label{PFwkexp}
\end{eqnarray}
\noindent
where
\begin{equation}
C(\beta ) = \exp\left[ 2\beta L^2 - \frac{3}{2}L^2\ln\beta \right] \ .
\label{preexp}
\end{equation}
\noindent
As usually, one has to put $h_k=s_k=0$ after taking all the derivatives. 
$M(h,s)$ is a generating functional which we study in the next subsection.
It is obvious that the ground state satisfies
\begin{equation}
<(\omega^{(0)}_k(x))^p> = 0  \ , \  p=1,2,...  \ ,
\label{mainst}
\end{equation}
\noindent
precisely like the abelian model.

\subsection{Generating functional and zero modes}

Here we are going to study the generating functional $M(h,s)$ given by
\begin{equation}
M(h,s) = \frac{Z(h,s)}{Z(0,0)} \ ,
\label{GF}
\end{equation}
\noindent
and
\begin{eqnarray}
Z(h,s) = \int_{-\infty}^{\infty} \prod_{x,k} d\alpha_k(x)
\int_{-\infty}^{\infty} \prod_{l,k} d\omega_k(l) 
\exp \left[ -\frac{1}{2}\omega^2_k(l) 
-i\omega_k(l)[\alpha_k(x+n) - \alpha_k(x) ] \right]  \nonumber  \\   
\sum_{m(x)=-\infty}^{\infty}
\exp \left[ 2\pi i\sqrt{2\beta}\sum_xm(x)\alpha (x) +
\sum_{l,k}\omega_k(l)h_k(l) + \sum_{x,k}\alpha_k(x)s_k(x) \right] .
\label{GF1}
\end{eqnarray}
\noindent
As in the abelian case we expect that integrals over zero modes
of the auxiliary field are not Gaussian and should lead to some constraint
on the sums over $m_x$. To see this, we put $h_k=s_k=0$ and integrate out
the link fields. Partition function becomes
\begin{equation}
Z(0,0) =\sum_{m(x)=-\infty}^{\infty} \int_{-\infty}^{\infty} \prod_{x,k} d\alpha_k(x)
\exp \left[ -\alpha_k(x)G_{x,x^{\prime}}^{-1}\alpha_k(x^{\prime})
+ 2\pi i\sqrt{2\beta}m(x)\alpha (x) \right] \ , 
\label{GF00}
\end{equation}
\noindent
with $G_{x,x^{\prime}}$ given in (\ref{Gxx}) and sum over repeating indices is
understood here and in what follows. 
It is clear that in this case the zero modes should be controlled via integration 
over the radial component of the vector $\vec{\alpha}(x)$. To see this, we change
to the spherical coordinates and treat only a constant mode in the angle variables
(since the zero mode problem could only arise from this configuration). One has
\begin{eqnarray}
Z(0,0) \sim \sum_{m_x=-\infty}^{\infty} \int_0^{\infty} 
\prod_x \alpha^2_x d\alpha_x
\exp \left[ -\alpha_xG_{x,x^{\prime}}^{-1}\alpha_{x^{\prime}}
+ 2\pi i\sqrt{2\beta}m_x\alpha_x \right]   \nonumber  \\
\sim \sum_{m_x=-\infty}^{\infty} \delta \left( \sum_xm_x \right) 
\exp \left[ -2\pi^2\beta \sum_{x,x^{\prime}}m_x
G_{x,x^{\prime}} m_{x^{\prime}}\right] + O(m_x^2) \ ,
\label{GF0mod}
\end{eqnarray}
\noindent
and we used the notation $\alpha_x$ for the radial component of the vector 
$\vec{\alpha}(x)$. Since the zero mode of the radial component in the $x$
space is
$$
\alpha (p=0) = \left( \frac{1}{L^D}\sum_k(\sum_x\alpha_k(x))^2 \right)^{1/2}
$$ 
one has to omit the zero mode from the Green function in each term 
of the sum over $k$ in the integrand of (\ref{GF00}).
As in the abelian case the integration over zero modes produces 
delta-function in (\ref{GF0mod}). Since however only $m_x=0$ for all $x$
contribute to the asymptotics of the free energy and fixed distance correlation
function (other values of $m_x$ being exponentially suppressed) 
we have to put $m_x=0$ omitting at the same time all zero modes. 
Calculating resulting Gaussian integrals we come to
\begin{equation}
M(h,s) = \exp 
\left[ \frac{1}{4}s_k(x)G_{x,x^{\prime}}s_k(x^{\prime}) +
\frac{i}{2}s_k(x)D_l(x)h_k(l) +
\frac{1}{4}h_k(l)G_{ll^{\prime}}h_k(l^{\prime}) \right] \ ,
\label{GFfin}
\end{equation}
\noindent
where $G_{ll^{\prime}}$ was introduced in (\ref{Gll1}) and
\begin{equation}
D_l(x^{\prime}) = G_{x,x^{\prime}} - G_{x+n,x^{\prime}} \ , \ l=(x,n) \ .
\label{Dxl}
\end{equation}
\noindent
From (\ref{GFfin}) one can deduce the following simple rules
\begin{eqnarray}
<\omega_k(l)\omega_n(l^{\prime})> =
\frac{\delta_{kn}}{2}G_{ll^{\prime}} \ , \ 
<\alpha_k(x)\alpha_n(x^{\prime})> =
\frac{\delta_{kn}}{2}G_{xx^{\prime}} \ ,  \nonumber  \\ 
- i<\omega_k(l)\alpha_n(x^{\prime})> =
\frac{\delta_{kn}}{2}D_l(x^{\prime}) \ .
\label{Frules}
\end{eqnarray}
\noindent
We describe some simple properties of the functions $G_{ll^{\prime}}$
and $D_l(x^{\prime})$ in the Appendix C.
The expansion (\ref{PFwkexp}), representation (\ref{GFfin}) for the 
generating functional and rules (\ref{Frules}) 
are main formulas of this section which allow 
to calculate the weak coupling expansion of both the free energy and any
short-distance observable. Let us now comment on the infrared
finitness of the expansion.
It follows from the representation for the generating functional
that all expectation values of the link fields are expressed only via
the link Green functions $G_{ll^{\prime}}$ and $D_l(x)$ which are infrared finite
by construction. All combinations of auxiliary fields which contain
odd overall powers of the fields are expressed only via $D_l(x)$
and, therefore are also infrared finite. However, even powers
include $G_{x,x^{\prime}}$ and the infrared finitness is not provided
automatically. In particular, it means that unlike $XY$ 
model we are not allowed to take the TL at this stage.

Our last comment concerns the partition function (\ref{GF00}). We believe it 
can be regarded as an analog of the corresponding expression in the $XY$ model,
i.e. this is a nonabelian analog of the so-called ``spin-wave--vortex'' representation
for the partition function. One can see that the nonabelian model
is not factorized into three abelian components and is periodic in
the length of the vector $\vec{\alpha}(x)$ rather than in 
its components $\alpha_k(x)$. 

\subsection{First order coefficient of the correlation function}

As the simplest example we would like to calculate the first
order coefficient of the correlation function (\ref{corf1}).
Expanding (\ref{corf1}) in $1/\beta$ one has
\begin{equation}
\Gamma (x,y) = 1 - \frac{1}{4\beta}
< \sum_{k=1}^3\left (\sum_l\omega_k(l) \right )^2 > + O(\beta^{-2}) = 
1 - \frac{3}{8\beta}
\sum_{l,l^{\prime}\in C^d_{xy}} G_{ll^{\prime}} + O(\beta^{-2}) \ ,
\label{GLXY1}
\end{equation}
\noindent
where $C^d_{xy}$ is a path dual to the path $C_{xy}$, i.e consisting
of the dual links which are orthogonal to the original
links $l,l^{\prime}\in C_{xy}$.
The form of $G_{ll^{\prime}}$ ensures independence of $\Gamma (x,y)$
of a choice of the path $C_{xy}$. After some algebra it is easy to get
the result
\begin{equation}
\Gamma (x,y) = 1 - \frac{3}{4\beta} D(x-y) \ , \
D(x) = \frac{1}{L^2} \sum_{k_n=0}^{L-1}
\frac{1 - e^{\frac{2\pi i}{L}k_n x_n}}
{D-\sum_{n=1}^D\cos \frac{2\pi}{L}k_n} \ , \ k_n^2\ne 0 ,
\label{Dx}
\end{equation}
\noindent
which coincides with the result of the conventional PT.

\section{First order coefficient of the free energy}

The main result of our study is the first order
coefficient of the $SU(2)$ free energy 
\begin{equation}
F=\frac{1}{2L^2}\ln Z = 2\beta - \frac{3}{4}\ln\beta +
\frac{1}{2\beta L^2}C^1 + O(\beta^{-2}) \ .
\label{fren}
\end{equation}
\noindent
There are four contributions at this order to $C^1$
\begin{equation}
C^1 = C^1_{ac} + C^1_{meas} + C^1_{J1} +  C^1_{J2} \ .
\label{frensum}
\end{equation}
\noindent
Contribution from the action (\ref{actexp}) is given by
\begin{equation}
\frac{1}{2L^2}C^1_{ac}=\frac{5}{128L^2}\sum_lG^2_{ll}=
\frac{5}{64} \ .
\label{c1ac}
\end{equation}
\noindent
Contribution from the measure (\ref{invMexp}) is given by
\begin{equation}
\frac{1}{2L^2}C^1_{meas}= - \frac{1}{8L^2}\sum_lG_{ll}
= - \frac{1}{4} \ .
\label{c1meas}
\end{equation}
\noindent
Contribution from third brackets in (\ref{PFwkexp})
is proportional to $[\omega_k^{(0)}]^2$ and equals 0,
because of (\ref{mainst}). There are two contributions
from the expansion of the Jacobian (\ref{J2exp}). The first one
is given by the expectation value of the operator
$-i\sum_x\sum_{k=1}^3\alpha_k(x)\omega_k^{(2)}(x)$.
$\omega_k^{(2)}(x)$ is given in the Appendix B (\ref{wkx2}). 
From the form of the generating functional (\ref{GFfin}) one can see that
the expectation value of this operator depends only on
link Green functions $G_{ll^{\prime}}$ and $D_l(x^{\prime})$.
One gets after long but straightforward algebra\footnote{Our previous
version suffered from incorrect sign in this term which led to 
wrong final result.}
\begin{eqnarray}
\frac{1}{2L^2}C^1_{J1}=\frac{1}{4L^2}\sum_l\sum_xD_l(x) \nonumber  \\
(\ \frac{1}{2}\sum_{i=1}^4(\delta_{ll_i}(G_{l_3l_i}+G_{l_4l_i}-
G_{l_1l_i}-G_{l_2l_i})+G_{l_il_i}(\delta_{ll_1}+\delta_{ll_2}-
\delta_{ll_3}-\delta_{ll_4}))+ \nonumber  \\
\delta_{ll_1}(G_{l_3l_4}+2G_{l_3l_2}+2G_{l_4l_2})+ 
\delta_{ll_2}(G_{l_3l_4}-G_{l_3l_1}-G_{l_4l_1})+ \nonumber  \\
\delta_{ll_3}(G_{l_4l_1}+G_{l_4l_2}-G_{l_1l_2})-
\delta_{ll_4}(2G_{l_3l_1}+2G_{l_3l_2}+G_{l_1l_2}) \ ) \ ,
\label{c1J1Gr}
\end{eqnarray}
\noindent
where links $l_i$ are defined in Appendix C (see Fig.1), $l=(x,n)$.
In terms of standard $D$-functions defined in Appendix C the result reads
\begin{equation}
\frac{1}{2L^2}C^1_{J1}=\frac{1}{4}[6-2D(2,0)-D(1,1)]=
\frac{1}{2}+\frac{3}{2\pi} \ .
\label{c1J1res}
\end{equation}
\noindent
The second term is given by the operator
$< \frac{1}{2}\left
(\sum_x\sum_{k=1}^3\alpha_k(x)\omega_k^{(1)}(x)\right )^2 >$.
$\omega_k^{(1)}(x)$ is given in the Appendix B (\ref{wkx1}).
In terms of Green functions it reads
\begin{equation}
\frac{1}{2L^2}C^1_{J2} = - \ \frac{3}{16} (Q^{(1)} + Q^{(2)}) \ ,
\label{c1J2res}
\end{equation}
\noindent
\begin{equation}
Q^{(1)}= \frac{1}{2L^2}\sum_{x,x^{\prime}}
\sum_{i<j}^4\sum_{i^{\prime}<j^{\prime}}^4
G_{x,x^{\prime}} 
(G_{l_i,l^{\prime}_{i^{\prime}}}G_{l_j,l^{\prime}_{j^{\prime}}} -
G_{l_i,l^{\prime}_{j^{\prime}}}G_{l_j,l^{\prime}_{i^{\prime}}}) \ ,
\label{Q1xx}
\end{equation}
\noindent
\begin{eqnarray}
Q^{(2)}=\frac{1}{2L^2}\sum_{x,x^{\prime}}
\sum_{i<j}^4\sum_{i^{\prime}<j^{\prime}}^4
(G_{l_i,l^{\prime}_{i^{\prime}}}D_{l^{\prime}_{j^{\prime}}}(x) 
D_{l_{j}}(x^{\prime}) + 
G_{l_j,l^{\prime}_{j^{\prime}}}D_{l^{\prime}_{i^{\prime}}}(x) 
D_{l_{i}}(x^{\prime}) -  \nonumber  \\
G_{l_i,l^{\prime}_{j^{\prime}}}D_{l^{\prime}_{i^{\prime}}}(x) 
D_{l_{j}}(x^{\prime}) - 
G_{l_j,l^{\prime}_{i^{\prime}}}D_{l^{\prime}_{j^{\prime}}}(x) 
D_{l_{i}}(x^{\prime})) \ . 
\label{Q2xx}
\end{eqnarray}
\noindent
Link $l_i$ ($l^{\prime}_{j^{\prime}}$) refers to one of four
links attached to a given site $x$ ($x^{\prime}$).
We performed both analytical and numerical studies of last expressions.
Details are given in the Appendix C. We find
\begin{equation}
Q^{(1)}= \frac{1}{2}
\label{Q1res}
\end{equation}
\noindent
and
\begin{equation}
Q^{(2)}= 1+\frac{8}{\pi} \ .
\label{Q2res}
\end{equation}
\noindent
Collecting all coefficients we finally obtain 
\begin{equation}
\frac{1}{2L^2}C^1 = \frac{3}{64} \ ,
\label{C1res}
\end{equation}
\noindent
which coincides with the result of the conventional PT
\cite{rossi}.

\section{Discussion}

In this paper we propose to use
an invariant link formulation to investigate some properties 
of $2D$ models in the weak coupling region. 
We argued that this approach is more suitable for calculation
of asymptotic expansions of invariant functions in cases
when the Mermin-Wagner theorem forbids spontaneous symmetry
breaking in the thermodynamic limit. 
We have found that both in the abelian $XY$ model and in non-abelian
$SU(2)$ model our results for the first order coefficients of 
the free energy and correlation function 
agree with the standard PT expansion in the TL.
In our next paper \cite{corrf} we show that the second
order coefficient of the correlation function also
agrees with the conventional PT. 

Now we can return to the question raised in the Introduction,
namely whether conventional PT gives uniform asymptotic
expansion for non-abelian models. It is well known for a long
time that it is no so in one-dimensional non-abelian models.
In \cite{corrf} we address this question in the link
formulation and show that non-uniformity in this case
originates from the expansion of holonomy operator
which imposes certain global condition on the configurations
of link matrices. While this global condition itself 
vanishes in true TL, it survives large volume limit if 
the low-temperature expansion is performed in a finite volume.
In $2D$ there are two such operators which we did not
consider in the present paper. The reason being
that the contributions from these holonomy operators
are suppressed as $O(1/L)$. Since in $2D$ one could 
encounter only logarithmic divergences it is rather
unlikely that holonomies are relevant for the TL.
In this sense there is no similarity between
one and two dimensional models. Nevertheless, 
one cannot exclude the possibility of non-uniformity
of the low-temperature expansion arising from
the remainder to the PT series \cite{QA}.
This problem is extremely hard to resolve by means
of the standard approaches, see for instance \cite{QUQ}.  
Contrary, in the link formulation we are able to
calculate the exponential remainder at a given order
of the low-temperature expansion. Therefore,
the problem of the infrared finitness of the remainder
can be addressed explicitly. Such investigations
are presently in progress.

\vspace{0.5cm} 
\begin{center}
{\bf Acknowledgements.} 
\end{center}

We are grateful to J.~Polonyi who found time to go
through many details of the expansion presented here
and for many encouraging discussions.
We would like to thank V.~Miransky and V.~Gusynin for interesting
discussions and healthy critics on different stages of this work.
Our special thanks to B.~Rusakov for the explanation of his
paper \cite{linkrepr} regarding the calculation of the Jacobian $J(V)$.

\section{Appendix}

\subsection{A: Representation for $SU(2)$ partition function}

We start from the following partition function in the link representation
\begin{equation}
Z = \int \prod_l dV_l \exp \left[ \beta \sum_l  {\mbox {Tr}} V_l \right] 
\prod_p J_p(V) \ .
\label{D1}
\end{equation}
\noindent
Using the formula for $SU(2)$ characters
\begin{equation}
\chi_r(\phi) = \frac{\sin (2r+1)\phi}{\sin \phi} \ ,
\label{D2}
\end{equation}
\noindent
with $r=0,1/2,1,3/2,...$,
we can write down the Jacobian $J(V)$ (\ref{jacob}) in the form
\begin{equation}
J_p(V) = \sum_{r_p=-\infty}^{\infty}r_p \frac{\sin r_p\phi_p}{\sin \phi_p} \ ,
\label{D3}
\end{equation}
\noindent
where $\phi_p$ is some plaquette angle. Now $r_p$ takes only integer values.
Let us parametrize the $SU(2)$ link matrices as
\begin{equation}
V_l = \exp \left [ i\sigma^k\omega_k(l) \right ] = \cos W_l + i\sigma^k\omega_k(l)
\frac{\sin W_l}{W_l} \ ,
\label{D4}
\end{equation}
\noindent
where we used $W_l$ defined in (\ref{Wl}).
One gets in this parametrization the following expressions:
\begin{enumerate}
\item Action
\begin{equation}
\frac{1}{2}{\mbox {Tr}} V_l = \cos W_l \ .
\label{D5}
\end{equation}
\noindent
\item Invariant measure
\begin{equation}
dV_l = \frac{\sin^2W_l}{W^2_l}\prod_k d\omega_k(l) \ .
\label{D6}
\end{equation}
\noindent
\item Jacobian
\begin{equation}
J_p(V) = \sum_{r_p=-\infty}^{\infty}r_p \frac{\sin r_pW_p}{\sin W_p} \ ,
\label{D7}
\end{equation}
\noindent
\end{enumerate}
\noindent
where $W_p$ is defined in (\ref{Wp}).
Substituting (\ref{D5})-(\ref{D7}) into (\ref{D1}) we get
\begin{equation}
Z = \int \prod_l \left[ \frac{\sin^2W_l}{W^2_l}\prod_kd\omega_k(l) \right]
\exp \left[ 2\beta\sum_l\cos W_l \right] \prod_p
\sum_{r_p=-\infty}^{\infty}r_p \frac{\sin r_pW_p}{\sin W_p} \ .
\label{D13}
\end{equation}
\noindent
To get an expression for the Jacobian convenient for the large-$\beta$
expansion we use the equality
\begin{equation}
\sum_{r=-\infty}^{\infty}r \frac{\sin rW}{\sin W} = \frac{W}{\sin W} 
\sum_{m=-\infty}^{\infty}\int\prod_kd\alpha_k 
\exp \left[ -i\sum_k\alpha_k\omega_k + 2\pi im\alpha \right] \ ,
\label{D14}
\end{equation}
\noindent
where we have introduced $\alpha = \left[ \sum_k\alpha^2_k \right]^{1/2}$.
To prove (\ref{D14}) we write
\begin{equation}
F(\sum_k\omega_k^2)=\int_{-\infty}^{\infty}\prod_kd\alpha_k
e^{-i\alpha_k\omega_k}\int_{-\infty}^{\infty}\prod_kdt_k
F(\sum_kt_k^2)e^{i\alpha_kt_k} \ .
\label{D16}
\end{equation}
\noindent
Integrals over $t_k$ are calculated in the spherical coordinates. 
Taking then $F(t^2)=\frac{\sin rt}{t}$ and using the Poisson resummation 
formula we come to (\ref{D14}).
Substituting (\ref{D14}) into the partition function (\ref{D13}) we finally get
\begin{eqnarray}
Z = \int \prod_l \left[ \frac{\sin^2W_l}{W^2_l}\prod_kd\omega_k(l) \right]
\exp \left[ 2\beta\sum_l\cos W_l \right] \prod_p \frac{W_p}{\sin W_p} \nonumber   \\
\prod_p \sum_{m(p)=-\infty}^{\infty}\int\prod_kd\alpha_k(p)
\exp \left[ -i\sum_k\alpha_k(p)\omega_k(p) + 2\pi im(p)\alpha (p) \right] \ .
\label{D19}
\end{eqnarray}
\noindent

\subsection{B: Relation between link and plaquette angles}

Let us introduce the following notations for links of a given plaquette $p$
\begin{equation}
l_1 = (x;n) \ , \ l_2 = (x+n;m) \ , \ l_3 = (x+m;n) \ , \ l_4 = (x;m) \ .
\label{D8}
\end{equation}
\noindent
Then, from the definition (\ref{plangle}) and (\ref{prod}) one gets
the following formulas relating link and plaquette angles
\begin{equation}
V_p = \cos W_p + i\sigma^k\omega_k(p) \frac{\sin W_p}{W_p} \ ,
\label{D9}
\end{equation}
\noindent
where
\begin{eqnarray}
\cos W_p = \cos M_1\cos M_2 + \sum_k\nu^k_1\nu^k_2 
\frac{\sin M_1\sin M_2}{M_1M_2} \ ,  \\
\omega_k(p) \frac{\sin W_p}{W_p} = \nu^k_1\cos M_2\frac{\sin M_1}{M_1} -
\nu^k_2\cos M_1\frac{\sin M_2}{M_2} + 
\epsilon^{kpq}\nu^p_1\nu^q_2 \frac{\sin M_1\sin M_2}{M_1M_2} \ .
\label{D10}
\end{eqnarray}
\noindent
$M_i$ and $\nu^k_i$ are given by
\begin{eqnarray}
\cos M_1 = \cos W(l_1)\cos W(l_2) - \sum_k\omega_k(l_1)\omega_k(l_2)
\frac{\sin W(l_1)\sin W(l_2)}{W(l_1)W(l_2)} \ ,  \\  \nonumber
\cos M_2 = \cos W(l_3)\cos W(l_4) - \sum_k\omega_k(l_3)\omega_k(l_4)
\frac{\sin W(l_3)\sin W(l_4)}{W(l_3)W(l_4)} \ ,  
\label{D11}
\end{eqnarray}
\noindent
\begin{eqnarray}
\nu^k_1\frac{\sin M_1}{M_1} =\omega_k(l_1)\cos W(l_2)\frac{\sin W(l_1)}{W(l_1)} +
\omega_k(l_2)\cos W(l_1)\frac{\sin W(l_2)}{W(l_2)} - \\ \nonumber
\epsilon^{kpq}\omega_p(l_1)\omega_q(l_2)
\frac{\sin W(l_1)\sin W(l_2)}{W(l_1)W(l_2)} \ , \\  \nonumber
\nu^k_2\frac{\sin M_2}{M_2} =\omega_k(l_3)\cos W(l_4)\frac{\sin W(l_3)}{W(l_3)} +
\omega_k(l_4)\cos W(l_3)\frac{\sin W(l_4)}{W(l_4)} -  \\  \nonumber
\epsilon^{kpq}\omega_p(l_4)\omega_q(l_3)
\frac{\sin W(l_3)\sin W(l_4)}{W(l_3)W(l_4)} \ .
\label{D12}
\end{eqnarray}
\noindent
Now it is straightforward to calculate the following expansion ($p\to x$)
\begin{equation}
\omega_k(x) =  \omega^{(0)}_k(x) +
\omega^{(1)}_k(x) + \omega^{(2)}_k(x) + ... \ .
\label{wkxexp}
\end{equation}
\noindent
On a dual lattice (see Fig.1) the first coefficients can be written down as
\begin{equation}
\omega^{(0)}_k(x) = \omega_k(l_3) + \omega_k(l_4) - \omega_k(l_1) - \omega_k(l_2) \ ,
\label{wkx0}
\end{equation}
\noindent
\begin{equation}
\omega^{(1)}_k(x) = -\epsilon^{kpq} \sum_{i<j}^4 \omega_p(l_i)\omega_q(l_j) \ ,
\label{wkx1}
\end{equation}
\noindent
\begin{eqnarray}
&&\omega^{(2)}_k(x)= \frac{1}{3}\epsilon^{mnc}\epsilon^{krc}\omega^{(0)}_n(x)
\sum_{i=1}^4 \omega_r(l_i)\omega_m(l_i)+ \nonumber  \\
&&\frac{2}{3}(\epsilon^{mnc}\epsilon^{krc}+\epsilon^{mrc}\epsilon^{knc})
[ \omega_r(l_3)\omega_m(l_1)\omega_n(l_2)+
\omega_r(l_4)\omega_m(l_1)\omega_n(l_2)-
\nonumber  \\
&&\omega_r(l_3)\omega_m(l_4)\omega_n(l_1)-
\omega_r(l_3)\omega_m(l_4)\omega_n(l_2) ] \ .
\label{wkx2}
\end{eqnarray}
\noindent

\subsection{C: Coefficients $Q^{(1)}$ and $Q^{(2)}$}

We define $D$-function as
\begin{equation}
D(x_1,x_2) = \frac{1}{L^2} \sum_{k_n=0}^{L-1}
\frac{1 - e^{\frac{2\pi i}{L}(k_1x_1+k_2x_2)}}
{2-\sum_{n=1}^2\cos \frac{2\pi}{L}k_n} \ , \ k_n^2\ne 0 ,
\label{Dx1x2}
\end{equation}
\noindent
and $G_0$ as
\begin{equation}
G_0 = \frac{1}{L^2} \sum_{k_n=0}^{L-1} \frac{1}
{2-\sum_{n=1}^2\cos \frac{2\pi}{L}k_n} \ , \ k_n^2\ne 0 .
\label{G0}
\end{equation}
\noindent
The normalization is chosen such that $G_{ll}=1$.
All $D$-functions can be expressed in terms of $D(1,0)$
and $D(1,1)$. In the TL one has
\begin{eqnarray}
D(1,0)=\frac{1}{2} \ , \ D(1,1)=\frac{2}{\pi} \ , \nonumber  \\
D(2,0)=2(2D(1,0)-D(1,1))=2-\frac{4}{\pi} \ ,  
D(2,1)=2D(1,1)-D(1,0)=\frac{4}{\pi}-\frac{1}{2} \ , \nonumber  \\
D(3,0)=4(D(2,0)-D(1,1))+D(1,0)=\frac{1}{2}+8(1-\frac{3}{\pi}) \ .
\label{Dvalues}
\end{eqnarray}
$G_0$ is known to diverge logarithmically in two dimensions
and behaves numerically as
\begin{equation}
G_0 = a_0 + a_1\ln L + o(1) \ , a_0\approx0.0974006 \ , \ 
a_1\approx0.318331\approx \frac{1}{\pi} \ .
\label{G0form}
\end{equation}
\noindent
We also need some representation for the function $G_{ll^{\prime}}$
in the momentum space. One finds from the definition (\ref{Gll1})
\begin{eqnarray}
G_{ll^{\prime}}=\frac{2\delta_{nn^{\prime}}-1}{L^2}
\sum_{k_n=0}^{L-1}\frac{e^{\frac{2\pi i}{L}k_n(x-x^{\prime})_n}}{f(k)}
F(n,n^{\prime}) \ , \nonumber  \\
F(n=n^{\prime})=2(1-\cos\frac{2\pi}{L}k_p) \ , n\ne p \ , \nonumber \\
F(n\ne n^{\prime})=(1-e^{\frac{2\pi i}{L}k_n})
(1-e^{-\frac{2\pi i}{L}k_{n^{\prime}}}) \ ,
\label{Gllmom}
\end{eqnarray}
where we have denoted
\begin{equation}
f(k)=2-\sum_{n=1}^2\cos \frac{2\pi}{L}k_n \ .
\label{fk}
\end{equation}
Using this representation it is easy to prove the following
``orthogonality'' relations for the functions $G_{ll^{\prime}}$
and $D_l(x)$ which are very useful for calculation of
lattice sums
\begin{eqnarray}
\sum_{b}G_{lb}G_{bl^{\prime}}=2G_{ll^{\prime}} \ ,   \\
\sum_{b}D_b(x)G_{bl^{\prime}}=0 \ , \\
\sum_{b}D_b(x)D_b(x^{\prime})=2G_{x,x^{\prime}} \ .
\label{orthog}
\end{eqnarray}

To visualize the summation over indices $i,i^{\prime}$ in 
(\ref{Q1xx}) and (\ref{Q2xx}) we depicted dual links in Fig.1.
\begin{figure}[t]
\centerline{\epsfxsize=4cm \epsfbox{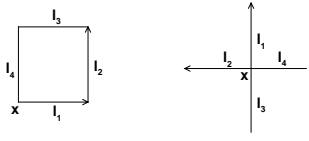}}
\hspace{0.5cm}
\caption{\label{dlink}Plaquette of original lattice and links of dual 
lattice as they enter sums for $Q^{(1)}$ and $Q^{(2)}$. 
Link is determined by point $x$ and a positive
direction, e.g. $l_3=(x-n_1;n_1)$.}
\end{figure}

\subsubsection{$Q^{(1)}$}

We divide $Q^{(1)}$ into two pieces
\begin{equation}
Q^{(1)} = B_1 + B_2 \ ,
\label{Q1B}
\end{equation}
where $B_1$ includes first and second powers of Green functions
$G_{x,x^{\prime}}$ whereas $B_2$ consists of terms with only
third powers of $G_{x,x^{\prime}}$. We thus have for $B_1$
from (\ref{Q1xx}) 
\begin{eqnarray}
B_1=\frac{1}{2L^2}\sum_{x,x^{\prime}}
\sum_{i<j}^4\sum_{i^{\prime}<j^{\prime}}^4
G_{x,x^{\prime}} 
(\delta_{l_i,l^{\prime}_{i^{\prime}}}G_{l_j,l^{\prime}_{j^{\prime}}} +
G_{l_i,l^{\prime}_{i^{\prime}}}\delta_{l_j,l^{\prime}_{j^{\prime}}} -
\delta_{l_i,l^{\prime}_{j^{\prime}}}G_{l_j,l^{\prime}_{i^{\prime}}} -
G_{l_i,l^{\prime}_{j^{\prime}}}\delta_{l_j,l^{\prime}_{i^{\prime}}} -
\nonumber   \\
\delta_{l_i,l^{\prime}_{i^{\prime}}}\delta_{l_j,l^{\prime}_{j^{\prime}}}+
\delta_{l_i,l^{\prime}_{j^{\prime}}}\delta_{l_j,l^{\prime}_{i^{\prime}}})  \ .
\label{B1def}
\end{eqnarray}
\noindent
Performing all summations and using Eq.(\ref{Dvalues}) to express
all $D$ functions appearing in the last equation in terms of
$D(1,0)$ and $D(1,1)$ we get
\begin{equation}
B_1 = 4D(1,1)-4D(1,0)-4G_0 \ .
\label{B1res}
\end{equation}
For $B_2$ we adduce an expression in the momentum space
\begin{equation}
B_2 = \frac{1}{2L^4}\sum_{k_n^1,k_n^2}\frac{B(k_n^1,k_n^2)}
{f(k^1)f(k^2)f(k^1+k^2)} \ , 
\label{B2mom}
\end{equation}
where
\begin{eqnarray}
B(k_n^1,k_n^2)=(z_1+z_2^*+z_1^*+z_2+z_2z_1^*+z_1z_2^*)
(z_4+z_3^*+z_3+z_4^*+z_4z_3^*+z_3z_4^*) + \\
2(z_1+z_1^*)(z_4+z_4^*) - 2z_1z_2^*z_4z_3^* - 2z_1z_2(z_3z_4)^* -
(z_1z_2+(z_1z_2)^*)(z_3z_4+(z_3z_4)^*) -  \nonumber  \\
z_1^2(z_3^*)^2-z_2^2(z_4^*)^2 -  
2z_1^*z_2^*(z_3^2+z_4^2) - 
2z_1z_2((z_3^*)^2+(z_4^*)^2) + 
(z_1^2-(z_1^*)^2)(z_4^2-(z_4^*)^2)  \nonumber  \\
\label{Bkk}  \nonumber  
\end{eqnarray}
and we have introduced
\begin{equation}
z_1=1-e^{ip_1^1} \ , \ z_2=1-e^{ip_2^1} \ , \ 
z_3=1-e^{ip_1^2} \ , \ z_4=1-e^{ip_2^2} \ ; \
p_n^i\equiv \frac{2\pi}{L}k_n^i \ .
\label{zi}
\end{equation}
The result of summation can be expressed in terms of $D$ functions
given in (\ref{Dvalues})
\begin{equation}
B_2 = 5D(1,0)-4D(1,1)+4G_0 \ .
\label{B2res}
\end{equation}
One sees that all divergences exactly cancel. The final result is
\begin{equation}
Q^{(1)} = B_1 + B_2=D(1,0)=\frac{1}{2} \ .
\label{Q1Bres}
\end{equation}
We performed numerical check of our result for $B_2$. 
The function was calculated for lattice sizes $L\in [10-120]$
and fitted to the form
\begin{equation}
B_2 = a_0+a_1\ln L + a_2\frac{\ln L}{L}+\frac{a_3}{L} \ .
\label{B2fit}
\end{equation}
From the result $Q^{(1)}=\frac{1}{2}$ and from Eq.(\ref{G0form})
one concludes that coefficients $a_0$ and $a_1$ should be equal to
$$
a_0=0.34312331 \ , \ a_1=1.273324 \ .
$$
Our fit gives
$$
a_0=0.343476 \ , \ a_1=1.27327 \ .
$$

\subsubsection{$Q^{(2)}$}

As before we divide $Q^{(2)}$ into two pieces
\begin{equation}
Q^{(2)} = K_1 + K_2 \ ,
\label{Q2K}
\end{equation}
where $K_1$ includes second powers of Green functions,
$K_2$ is cubic in $G_{x,x^{\prime}}$. $K_1$ is given by
\begin{eqnarray}
K_1=\frac{1}{2L^2}\sum_{x,x^{\prime}}
\sum_{i<j}^4\sum_{i^{\prime}<j^{\prime}}^4
(\delta_{l_i,l^{\prime}_{i^{\prime}}}D_{l^{\prime}_{j^{\prime}}}(x) 
D_{l_{j}}(x^{\prime}) + 
\delta_{l_j,l^{\prime}_{j^{\prime}}}D_{l^{\prime}_{i^{\prime}}}(x) 
D_{l_{i}}(x^{\prime}) -  \nonumber  \\
\delta_{l_i,l^{\prime}_{j^{\prime}}}D_{l^{\prime}_{i^{\prime}}}(x) 
D_{l_{j}}(x^{\prime}) - 
\delta_{l_j,l^{\prime}_{i^{\prime}}}D_{l^{\prime}_{j^{\prime}}}(x) 
D_{l_{i}}(x^{\prime})) \ . 
\label{K1}
\end{eqnarray}
Calculations are rather lengthy nevertheless straightforward.
We get
\begin{equation}
K_1=20D(1,0)^2+4(D(2,0)-D(1,0))(2D(1,1)-3D(1,0)+D(2,0))=
8-\frac{8}{\pi} \ .
\label{K1res}
\end{equation}
$K_2$ in the momentum space can be written as
\begin{equation}
K_2 = \frac{1}{L^4}\sum_{k_n^1,k_n^2}
\frac{K(k_n^1,k_n^2)}{f(k^1)f(k^2)f(k^1+k^2)} \ . 
\label{K2mom}
\end{equation}
where
\begin{eqnarray}
K(k_n^1,k_n^2)=
(z_1+z_2+z_3+z_4-z_1z_3-z_2z_4)[(z_1+z_2)^2(z_3^*+z_4^*) 
\nonumber  \\
- (z_1+z_1^*+z_2+z_2^*)(z_3+z_4)] \nonumber  \\
-2z_1z_4^*[2(z_2+1)(z_2^*+z_4^*-z_2^*z_4^*)+
z_1(z_2+z_4-z_2z_4)] \ .
\label{Kkk}
\end{eqnarray}
Notations are as in Eq.(\ref{zi}).
All sums can be done analytically resulting in\footnote{This analytical 
expression agrees well with our previous numerical result
$Q^{(2)}=3.5466309$}
\begin{equation}
K_2 = \frac{16}{\pi} -7 \ .
\label{K2res}
\end{equation}
Finally we obtain
\begin{equation}
Q^{(2)} = \frac{8}{\pi} + 1\approx 3.5464791 \ . 
\label{Q2Kres}
\end{equation}

\end{document}